\documentclass[twocolumn,showpacs,preprintnumbers,amsmath,amssymb]{revtex4}

\usepackage{graphicx}
\usepackage{dcolumn}
\usepackage{bm}

\begin{document}

\title{Isospin dependence of projectile-like fragment production at intermediate energies}
\author{C. W. Ma}\email{machunwang@126.com}
 \affiliation{College of Physics and Information Engineering, Henan Normal University, Xinxiang 453007, Peoples's Republic of China}
\author{H. L. Wei}
 \affiliation{College of Physics and Information Engineering, Henan Normal University, Xinxiang 453007, Peoples's Republic of China}
\author{Y. Fu}
 \affiliation{ Shanghai Institute of Applied Physics, Chinese Academy of Sciences,\\
P.O. Box 800-204, Shanghai 201800, Peoples's Republic of China}
\author{D. Q. Fang}
 \affiliation{ Shanghai Institute of Applied Physics, Chinese Academy of Sciences,\\
P.O. Box 800-204, Shanghai 201800, Peoples's Republic of China}
\author{W. D. Tian}
 \affiliation{ Shanghai Institute of Applied Physics, Chinese Academy of Sciences,\\
P.O. Box 800-204, Shanghai 201800, Peoples's Republic of China}
\author{X. Z. Cai}
 \affiliation{ Shanghai Institute of Applied Physics, Chinese Academy of Sciences,\\
P.O. Box 800-204, Shanghai 201800, Peoples's Republic of China}
\author{H. W. Wang}
 \affiliation{ Shanghai Institute of Applied Physics, Chinese Academy of Sciences,\\
P.O. Box 800-204, Shanghai 201800, Peoples's Republic of China}
\author{Y. G. Ma}
 \affiliation{ Shanghai Institute of Applied Physics, Chinese Academy of Sciences,\\
P.O. Box 800-204, Shanghai 201800, Peoples's Republic of China}
\author{J. Y. Wang}
 \affiliation{College of Physics and Information Engineering, Henan Normal University, Xinxiang 453007, Peoples's Republic of China}
\author{G. J. Liu}
 \affiliation{College of Physics and Information Engineering, Henan Normal University, Xinxiang 453007, Peoples's Republic of China}

\date{\today}

\begin{abstract}
The cross sections of fragments produced in 140 $A$ MeV $^{40,48}$Ca
+ $^9$Be and $^{58,64}$Ni + $^9$Be reactions are calculated by the
statistical abration-ablation(SAA) model and compared to the
experimental results measured at the National Superconducting
Cyclotron Laboratory (NSCL) at Michigan State University. The
fragment isotopic and isotonic cross section distributions of
$^{40}$Ca and $^{48}$Ca, $^{58}$Ni and $^{64}$Ni, $^{40}$Ca and
$^{58}$Ni, and $^{48}$Ca and $^{64}$Ni are compared and the isospin
dependence of the projectile fragmentation is studied. It is found
that the isospin dependence decreases and disappears in the central
collisions. The shapes of the fragment isotopic and isotonic cross
section distributions are found to be very similar for symmetric
projectile nuclei. The shapes of the fragment isotopic and isotonic
distributions of different asymmetric projectiles produced in
peripheral reactions are found very similar. The similarity of the
distributions are related to the similar proton and neutron density
distributions inside the nucleus in framework of the SAA model.
\end{abstract}

\pacs{25.70.Mn, 21.65.Cd}
\maketitle
\section{Introduction}
Projectile fragmentation is a well-established technique for the
production of rare isotope beams used by many radioactive ion-beam
facilities around the world. The process of projectile fragmentation
has been studied extensively to investigate the reaction mechanisms
in heavy ion collisions at intermediate and high energies.
Understanding the physics of projectile fragmentation is important
not only for rare-isotope beam production purposes but also for the
fundamental nuclear physics processes involved in nuclear collisions
\cite{Bao96,Brown00,Tsan01,YGMplb,YGM05,HAU98,Mayg07}.

Isospin effect is the phenomenon induced by the isospin degree of
freedom in heavy-ion collisions. Isospin effects of various physical
phenomena, such as multifragmentation, collective flow,
preequilibrium nucleon emission, etc., have been reported
\cite{MIL99,Demp96,Bao96,Pak971,Pak972,Chen98,Kum98,Ma00act}. These
studies have shown that isospin effect exists in nuclear reactions
induced by exotic nuclei but it may disappear under some conditions.
In projectile fragmentation reactions, the yields of neutron-rich
nuclei from fragmentation of neutron-rich projectiles will be larger
than those from stable nuclei; this is one isospin effect in
fragment production\cite{MIL99,FANG00,FANG01,FANG00cpl}. But the
difference becomes smaller with the increase of the charge
difference between the fragment and the projectile. And it may
disappear at last. In Refs. \cite{FANG00} and \cite{FANG00cpl}, the
fragment isotopic distributions of 60 $A$ MeV $^{16,18}$O + $^9$Be,
$^{36,40}$Ar + $^9$Be and $^{40,48}$Ca + $^9$Be were presented by
Fang $et~al$. It was found that the peak position of isotopic
distributions from stable nucleus induced reactions has a shift
toward the neutron-rich side as compared to that from neutron-rich
nucleus induced reactions. The shift becomes larger when the
difference between neutron numbers of the two projectiles is bigger.
This phenomena is called the isospin effect in the projectile
fragmentation. But the isospin effect of fragmentation reaction on
the isotopic distribution decreases with the increase of the atomic
number difference and disappears at last.

\begin{figure*}[tbp]
\includegraphics[width=16.5cm]{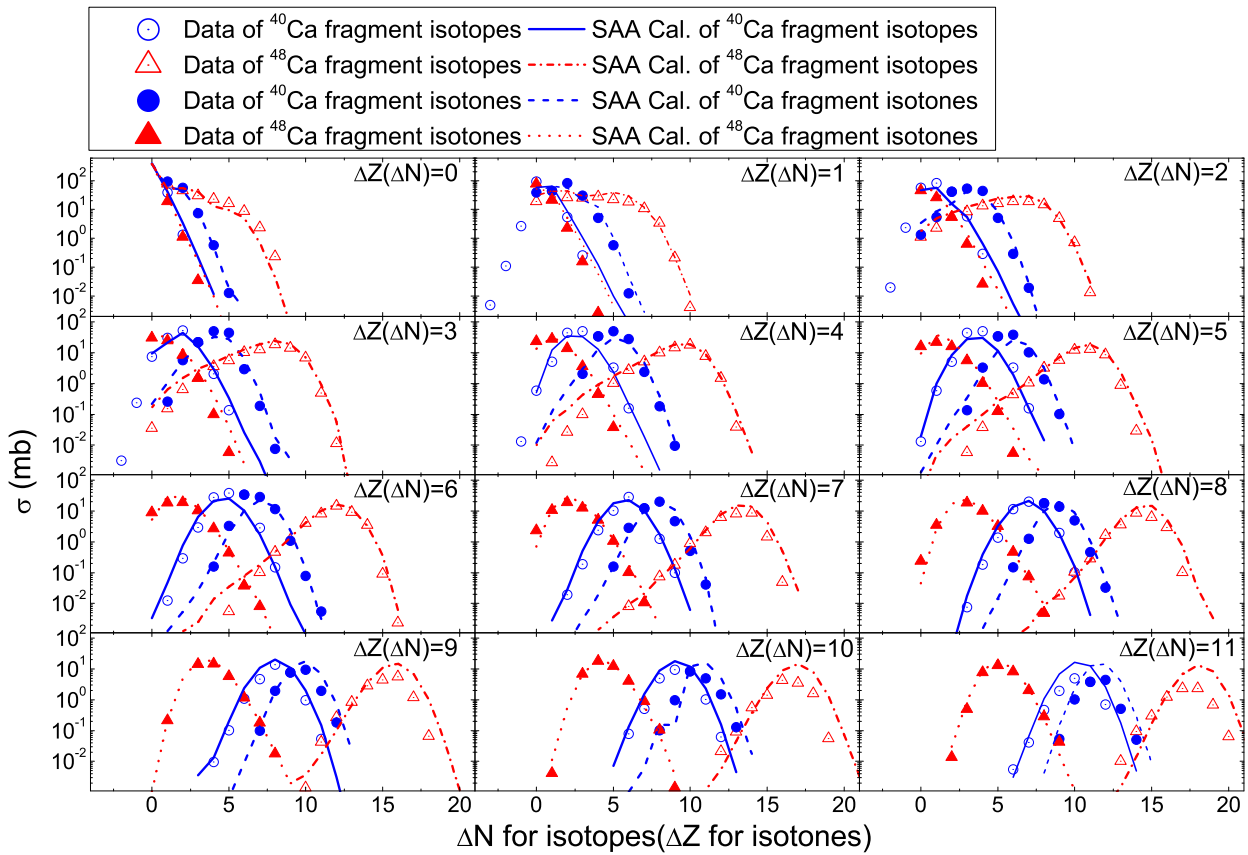}
\caption{(Color online) The cross sections of fragments produced in
the 140 $A$ MeV $^{40,48}$Ca + $^9$Be reactions. The open circles
denote the fragment isotopic $\sigma$ distributions, and the filled
ones denote the fragment isotonic $\sigma$ distributions of the
$^{40}$Ca + $^9$Be reaction. The $x$ axis represents the numbers of
protons removed($\Delta$Z = Z$_{\mbox{proj}}~-$ Z$_{\mbox{frag}}$)
from the projectile for the isotones, and the numbers of neutrons
removed($\Delta$N = N$_{\mbox{proj}}~-$ N$_{\mbox{frag}}$) from the
projectile for the isotopes in the pad. The open triangles denote
the fragment isotopic $\sigma$ distributions and the solid triangles
denote the fragment isotonic $\sigma$ distributions of $^{48}$Ca +
$^9$Be reaction. The lines are the results of the SAA model
calculations.} \label{Cafragdist}
\end{figure*}

In the campaign of four projectile fragmentation experiments carried
out at the National Superconducting Cyclotron Laboratory (NSCL) at
Michigan State University\cite{Mocko06}, eight different reaction
systems and more than 1400 fragment cross sections have been
measured. Reactions of primary 140 $A$ MeV of $^{40}$Ca, $^{48}$Ca,
$^{58}$Ni, and $^{64}$Ni colliding with light target $^9$Be and
heavy target $^{181}$Ta were measured. Extensive study of the
projectile fragmentation reactions using the \textbf{EPAX}
code\cite{Mocko06}, the macroscopic-microscopic heavy ion phase
space exploration (HIPSE) model, and the fully microscopic
antisymmetrized molecular dynamics (AMD) model has been carried
out\cite{Mocko08}. With no variation of the model parameters, a
reasonable agreement between the predictions and experimental data
has been reached\cite{Mocko08}. Comprehensive fragment cross
sections of $^{40}$Ca, $^{48}$Ca, $^{58}$Ni and $^{64}$Ni were
presented in Refs. \cite{Mocko06} and \cite{Mocko08}. From the
detailed cross section of fragments presented, it is possible to
investigate the isospin effect of projectile fragmentation induced
by the symmetric and asymmetric (neutron-rich) nuclei.

One of the methods used to calculate the cross section of fragments
produced in projectile fragmentation is the statistical abrasion
ablation (SAA) model. Compared to the HIPSE model and the AMD model,
the calculation of the SAA model is simple and can reproduce the
experimental results of heavy ion collision at intermediate energy
well\cite{FANG00,FANG01,FANG99,ZHC03,ZHC06}. In this article, the
cross section of fragments produced in 140 $A$ MeV $^{40,48}$Ca +
$^9$Be and $^{58,64}$Ni + $^9$Be reactions are calculated within the
framework of the SAA model and compared to the NSCL experimental
data \cite{Mocko08}. The fragment isotopic and isotonic cross
section distributions between the four reactions is compared to
study the isospin dependence of projectile-like fragmentation for
symmetric and asymmetric nuclei.

\section{The SAA Model}

The SAA model was developed by Brohm and Schmidt to describe the
peripheral nuclear collisions at high energies in a picture of
quasi-free nucleon-nucleon collisions\cite{Brohm94}. It was modified
by Fang and Zhong \textit{et~al.} to study the heavy ion collisions
at intermediate energy\cite{FANG00,FANG01,FANG99,ZHC03,ZHC06,MA08}.
The reactions in the SAA model are described as a two-step process.
The initial stage can be described by a Glauber-type model as
"participants" and "spectators." The participants in an overlapping
region between the projectile and the target interact strongly while
the spectators are left to move almost without being
disturbed\cite{EIS54}. In the second evaporation stage, the system
reorganizes because of excitation, which means that it is deexcited
and thermalized by the cascade evaporation of light particles. After
the deexcitation, the results of the final fragment, which are
comparable to the experimental data, can be obtained. The details of
the SAA model can be found in Refs. \cite{Brohm94,FANG00,FANG99}.

\begin{figure*}[tbp]
\includegraphics[width=16.5cm]{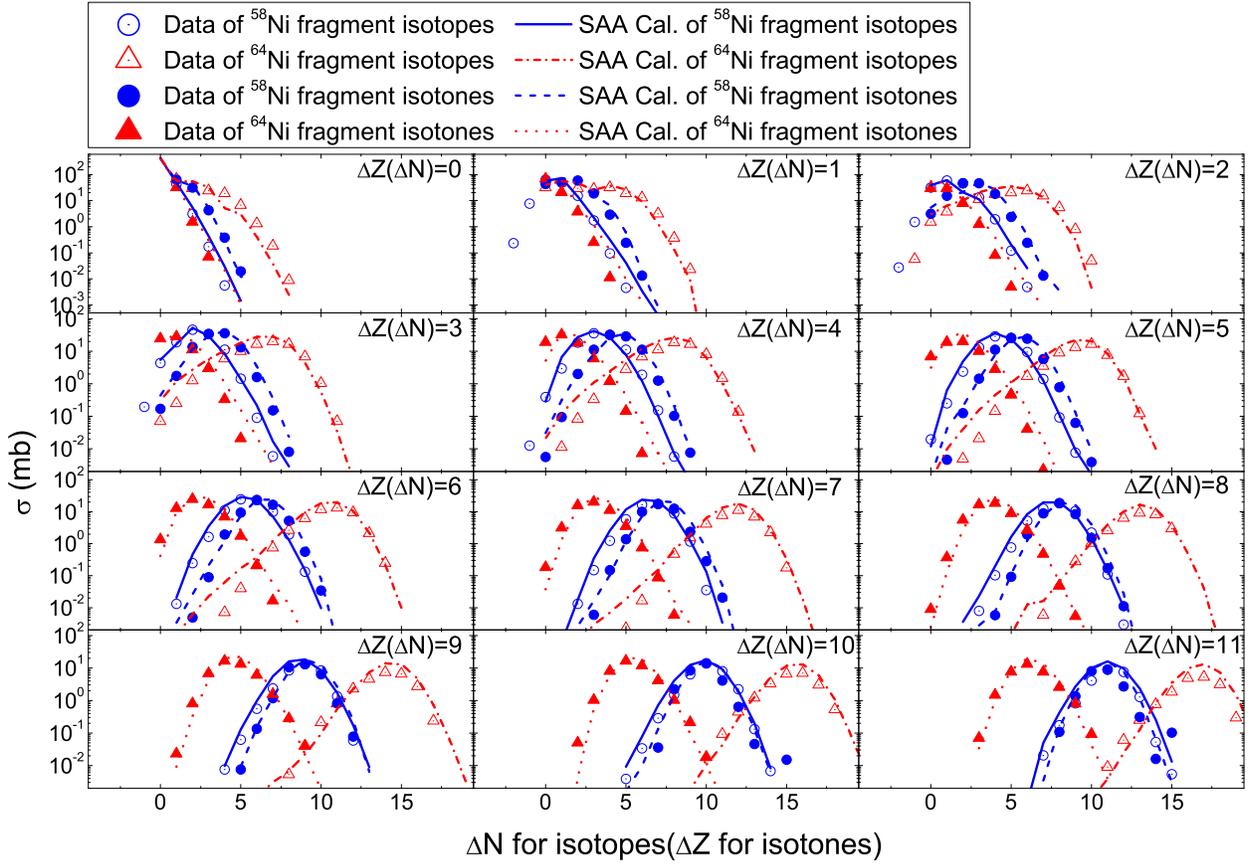}
\caption{(Color online) The cross sections of fragments produced in
the 140 $A$ MeV $^{58,64}$Ni + $^9$Be reactions. The open circles
denote the fragment isotopic $\sigma$ distributions, and the solid
ones denote the fragment isotonic $\sigma$ distributions of the
$^{58}$Ni+$^9$Be reaction. The $x$ axis represents the numbers of
protons removed($\Delta$Z = Z$_{\mbox{proj}} -$ Z$_{\mbox{frag}}$)
from the projectile for the isotones, and the numbers of neutrons
removed($\Delta$N=N$_{\mbox{proj}} -$ N$_{\mbox{frag}}$) from the
projectile for the isotopes in the pad. The open triangles denote
the fragment isotopic $\sigma$ distributions and the solid triangles
denote the fragment isotonic $\sigma$ distributions of the $^{64}$Ni
+ $^9$Be reaction. The lines denote the results of the SAA model
calculations.} \label{Nifragdist}
\end{figure*}

In the SAA model, the colliding nuclei are described to be composed
of parallel tubes orienting along the beam direction. Neglecting the
transverse motion, the collision is described by independent
interactions of tube pairs. Assuming a binomial distribution for the
absorbed projectile neutrons and protons in the interaction of a
specific pair of tubes, the distributions of the total abraded
neutrons and protons can be determined. For an infinitesimal tube in
the projectile, the transmission probabilities for neutrons
(protons) at a given impact parameter $\mathit{\textbf{b}}$ are
calculated by
\begin{equation}
t_k(\mathbf{s}-\mathbf{b})=\mbox{exp}\{-[D{_n^T}(\mathbf{s}-\mathbf{b})\sigma_{nk}+D{_n^P}(\mathbf{s}-\mathbf{b})\sigma_{pk}]\}
\end{equation}
where $D^T$ is the nuclear-density distribution of the target
intergrated along the beam direction and normalized by $\int
d^{2}sD_{n}^{T}=N^T$ and $\int d^{2}sD_{p}^{T}=Z^T$. $N^T$ and $Z^T$
refer to the neutron and proton numbers of the target, respectively.
The vectors $\mathit{\textbf{s}}$ and $\mathit{\textbf{b}}$ are
defined in the plane perpendicular to the beam. $\sigma_{k'k}$ is
the free space nucleon-nucleon cross section ($k',k=n$ for neutron
and $k',k=p$ for proton). The average absorbed mass in the limit to
infinitesimal tubes at a given $\mathit{\textbf{b}}$ is
\begin{eqnarray}
<\Delta A(b)>=\int d^{2}sD_{n}^{T}(\mathbf{s})[1-t_n(\mathbf{s}-\mathbf{b})] \nonumber\\
+\int d^{2}sD_{p}^{P}(\mathbf{s})[1-t_p(\mathbf{s}-\mathbf{b})]
\end{eqnarray}
The excitation energy of projectile spectator is estimated by the
simple relation of $E^*=13.3<\Delta A(b)>$ MeV, where 13.3 is the
mean excitation energy due to an abraded nucleon from the initial
projectile\cite{GAIM91}.

The production cross section for a specific isotope can be
calculated from
\begin{equation}
\sigma(\Delta N, \Delta Z)=\int d^2bP(\Delta N, b)P(\Delta Z,b),
\end{equation}
where $P(\Delta N,b)$ and $P(\Delta Z, b)$ are the probability
distributions for the abraded neutrons and protons at a given impact
parameter $\mathit{b}$, respectively.

The $\sigma$ of fragments produced in 140 $A$ MeV $^{40,48}$Ca +
$^9$Be and $^{58,64}$Ni + $^9$Be reactions are calculated within the
framework of the SAA model. In the calculations, the free space
nucleon-nucleon cross sections are adopted\cite{Cai98}. The proton
and neutron density distributions are assumed to be the Fermi type,
\begin{equation}
\rho_i(r)=\frac{\rho_i^0}{1+\mbox{exp}(\frac{r-C_i}{t_i/4.4})},~~~
i=n,p
\end{equation}
where $\rho_i^0$ is the normalization constant that ensures that the
integration of the density distribution equals the number of
neutrons ($i=n$) or protons ($i=p$); $t_i$ is the diffuseness
parameter, and $C_i$ is half the density radius of the neutron or
proton density distribution.

\section{Results and discussion}

\begin{figure*}[tbp]
\includegraphics[width=16.5cm]{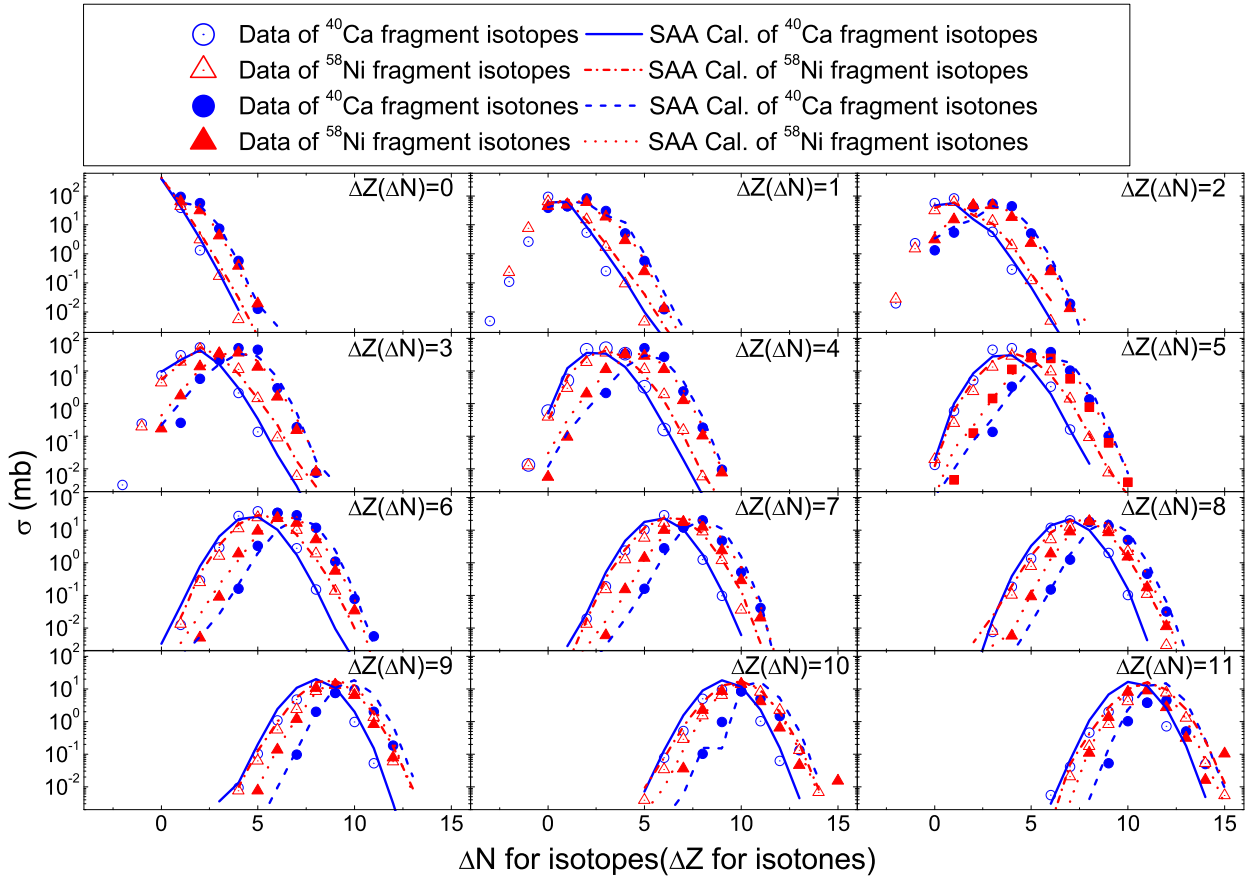}
\caption{(Color online) The cross sections of fragments produced in
the 140 $A$ MeV $^{40}$Ca + $^9$Be and $^{58}$Ni + $^9$Be reactions.
The open circles denote the fragment isotopic $\sigma$
distributions, and the solid ones denote the fragment isotonic
$\sigma$ distributions of the $^{40}$Ca + $^9$Be reaction. The open
triangles denote the fragment isotopic $\sigma$ distributions and
the solid triangles denote the fragment $\sigma$ distributions of
the $^{58}$Ni+$^9$Be reaction. The lines denote the results of the
SAA model calculations.} \label{Ca40Ni58}
\end{figure*}

In Fig.~\ref{Cafragdist}, the $\sigma$ of fragments produced in 140
$A$ MeV $^{40,48}$Ca + $^9$Be are plotted. In Fig.~\ref{Nifragdist},
the $\sigma$ of fragments produced in $^{58,64}$Ni+$^9$Be reactions
are plotted. The $x$ axis is not the usually used mass number but
the number of removed neutrons($\Delta
N=N_{\mbox{proj}}-N_{\mbox{frag}}$) for isotopes or the number of
removed protons($\Delta Z=Z_{\mbox{proj}}-Z_{\mbox{frag}}$) for
isotones from the projectile, which not only can reflect the fine
information of the reactions, but also makes it convenient to
compare between the fragment isotopic and isotonic distributions of
different projectiles. From Figs.~\ref{Cafragdist} and
\ref{Nifragdist}, it can be seen that the SAA model reproduces the
experimental data quite well not only for the stable nucleus
projectile but also for the neutron-rich nucleus projectile.

In Refs. \cite{FANG00} and \cite{FANG00cpl}, the fragment isotopic
distributions of different projectiles are compared. The isospin
effect and its disappearance in the projectile fragmentation is
found. There is a shift from the fragment isotopic $\sigma$
distributions (fragment isotopic distribution) of $^{40}$Ca to that
of $^{48}$Ca. There is also a shift from the isotonic $\sigma$
distributions of $^{40}$Ca to that of $^{48}$Ca. It is the same
phenomena as observed in Refs. \cite{FANG00} and \cite{FANG00cpl};
i.e., for the neutron-rich projectile, neutrons are more easily
removed than for the symmetric nucleus projectile.

\begin{figure*}[tbp]
\includegraphics[width=16.5cm]{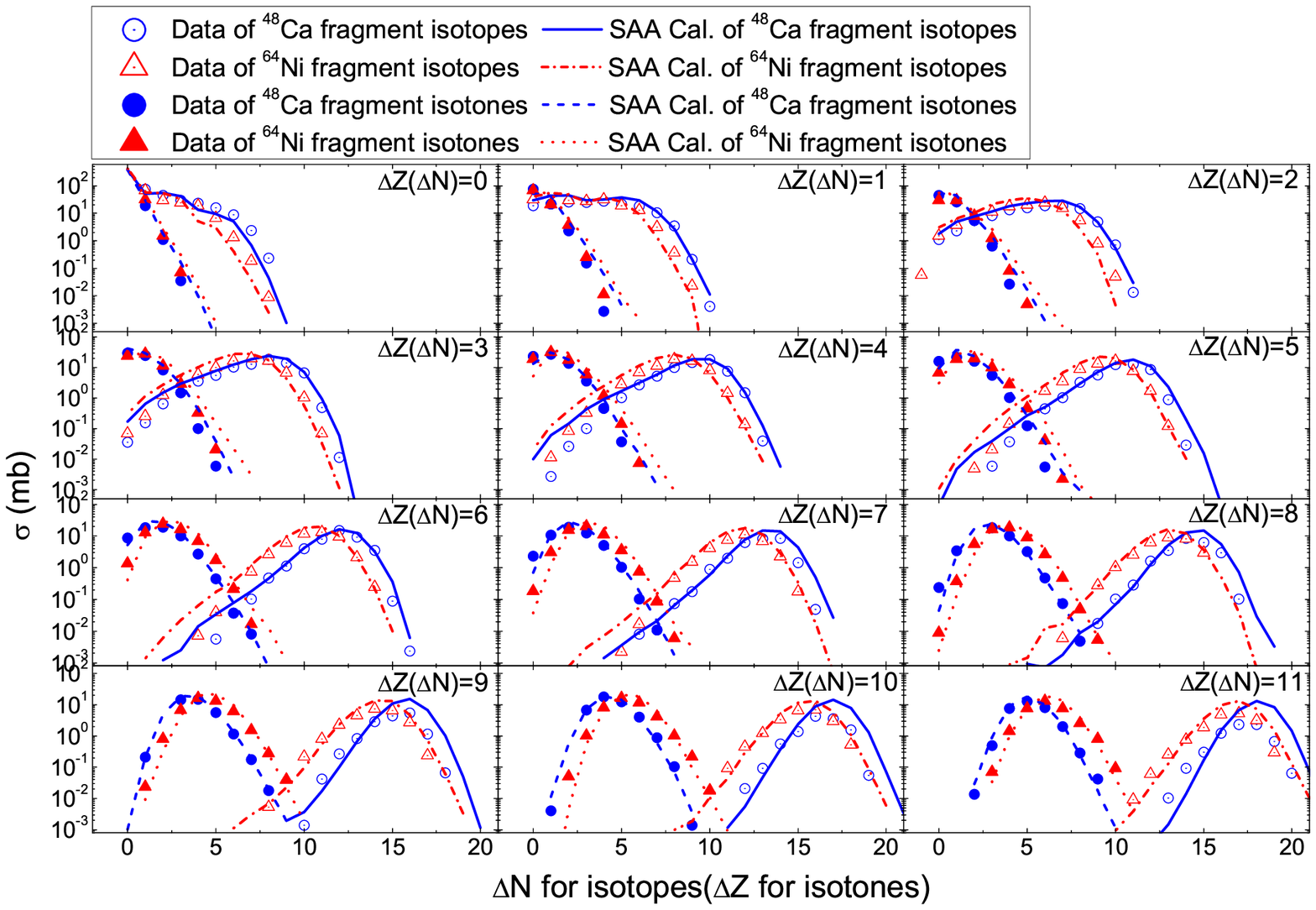}
\caption{(Color online) The cross sections of fragments produced in
the 140 $A$ MeV $^{48}$Ca + $^9$Be and $^{64}$Ni + $^9$Be reactions.
The open circles denote the fragment isotopic $\sigma$
distributions, and the solid ones denote the fragment isotonic
$\sigma$ distributions of the $^{48}$Ca+$^9$Be reaction. The open
triangles denote the fragment isotopic $\sigma$ distributions and
the solid triangles denote the fragment isotonic $\sigma$
distributions of the $^{64}$Ni + $^9$Be reaction. The lines denote
the results of the SAA model calculations.} \label{Ca48Ni64}
\end{figure*}

For $^{40}$Ca, in Fig.~\ref{Cafragdist}, the shapes of the fragment
isotopic and isotonic distributions are similar but there is a shift
from the isotopic to the isotonic distributions. The shift indicates
the isospin dependence of the fragment production. The shift becomes
smaller when $\Delta Z$ of fragment isotopes and $\Delta N$ of
fragment isotones increases, which means that central collisions
begin to dominate the reactions. In other words, the isospin effect
decreases in central collisions. The fragment isotopic distributions
of $^{48}$Ca are wider than those of $^{40}$Ca. It is easy to
understand because $^{48}$Ca is a neutron-rich nucleus and has a
neutron skin structure, from which neutrons can be removed more
easily. The shift between the fragment isotopic and isotonic
distributions of $^{48}$Ca is in the reverse direction of that of
$^{40}$Ca and become wider compared to those of $^{40}$Ca, which
indicates that the isospin effect in $^{48}$Ca projectile
fragmentation does not decrease but is enhanced in central
collisions.

The shapes of fragment isotopic and isotonic distributions of
$^{58}$Ni are similar to those of $^{40}$Ca. The fragment isotopic
and isotonic distributions of $^{64}$Ni are similar to those of
$^{48}$Ca. What interested us most in Fig.~\ref{Nifragdist} is that
the fragment isotopic and isotonic distributions overlap when
$\Delta Z$ of isotopes and $\Delta N$ of isotones are more than 9.
It is the direct evidence of the disappearance of the isospin
dependence of the projectile fragmentation. Compared to that of
$^{40}$Ca, it also means there is less isospin dependence in the
projectile fragmentation of $^{58}$Ni.

From the investigations of the fragment isotopic and isotonic
distributions of $^{40}$Ca and $^{58}$Ni, evidence for the decrease
and disappearance of the isospin effect in projectile fragmentation
is found; i.e., the shift from the fragment isotopic distributions
to the fragment isotonic distributions of $^{40}$Ca and $^{58}$Ni
becomes narrower and the fragment isotopic and isotonic
distributions of $^{58}$Ni overlap.

There is little difference between the proton and neutron density
distributions inside the symmetric nucleus. For $^{40}$Ca and
$^{58}$Ni, with $N/Z$ equal to 1.0 and 1.071, respectively, the
proton and neutron density distributions should be very similar
except the size of $^{58}$Ni is larger than that of $^{40}$Ca. There
are some reasons to expect similar fragment isotopic and isotonic
distributions of their projectile fragmentation. In Fig.~
\ref{Ca40Ni58}, the $\sigma$ values of fragments produced in
$^{40}$Ca/$^{58}$Ni + $^9$Be reactions are plotted. Though the
fragment isotopic distribution of $^{40}$Ca is narrower than that of
$^{58}$Ni, the difference between the fragment isotopic
distributions of $^{40}$Ca and $^{58}$Ni is small. The same behavior
is exhibited in the fragment isotonic distributions. Because the
difference between $^{40}$Ca and $^{58}$Ni mass numbers is not very
big, it can be concluded that, for symmetric projectile nuclei with
similar mass numbers, the fragment isotopic and isotonic
distributions should be similar. Studying these distributions more
carefully, it can be seen that the $\sigma$ values of fragments
produced in peripheral reactions of $^{40}$Ca and $^{58}$Ni have
less difference than those in central collisions. For the $\Delta
Z\geq3$ isotopes, the $\sigma$ values of fragments of $^{58}$Ni are
higher than those of $^{40}$Ca on the right sides of the
distributions while the left sides of the distributions show very
little difference. For the $\Delta N\geq3$ isotones, the $\sigma$
values of fragments of $^{58}$Ni are higher than those of $^{40}$Ca
on the left sides of the distributions while the right sides of the
distributions show very little difference.

In Fig.~\ref{Ca48Ni64}, the $\sigma$ values of fragments produced in
$^{48}$Ca /$^{64}$Ni + $^9$Be reactions are plotted. There is very
little difference between the fragment isotopic distributions of
$^{48}$Ca and $^{64}$Ni in peripheral reactions. For the $\Delta
Z<3$ isotopes, though the right sides (more nucleons removed) of the
isotopic distributions have a shift from the fragments of $^{48}$Ca
to $^{64}$Ni, the left sides (little nucleons removed) of the
isotopic distributions show very little difference. The $\Delta N<3$
isotones of $^{48}$Ca and $^{64}$Ni overlap. The fragments on the
left side of the $\Delta Z<3$ isotopes and the $\Delta N<3$ isotones
are the productions of most peripheral reactions. In Fig.
\ref{Ca48Ni64}, the shapes of the $\Delta Z\geq3$ isotopic
distributions of $^{48}$Ca and $^{64}$Ni are very similar but the
isotopic distributions of $^{48}$Ca shift to those of $^{64}$Ni and
the right sides of the distributions overlap gradually. Compared to
the left side of the distribution, the right side of the
distribution means that the central collisions happen most. For
$^{48}$Ca and $^{64}$Ni, the distributions of $\Delta Z\leq5$
fragment isotones, which are the productions in peripheral
reactions, show very little difference.

By investigating the fragment distributions of the four projectiles,
the isospin effect in the projectile fragmentation is exhibited. By
comparing the fragment isotopic and isotonic distributions of
$^{40}$Ca and $^{58}$Ni, it has been found that the isospin
dependence of the projectile fragmentation decreases and disappears.
The similarities of the fragment isotopic and isotonic distributions
from projectile fragmentation are discovered not only in symmetric
nuclei but also in asymmetric nuclei. It is meaningful to discover
the similarities of the fragment isotopic and isotonic distributions
because it help us to estimate the production of fragments in
projectile fragmentation at intermediate energy. But why do these
phenomena happen?

According to the SAA model, the number of protons or neutrons
removed in a specific pipe in collision is determined by the
nucleon-nucleon cross section and the densities of the protons and
neutrons. Because the projectile fragmentation of $^{40,48}$Ca and
$^{58,64}$Ni occur under the same experimental conditions, the
deexcitation in the reactions is the same and there is no need to
consider the target effect in the reactions. Some qualitative
explanations of the phenomena observed above can be obtained from
the SAA model.

For symmetric nuclei, as have been referred, their proton and
neutron density distributions are similar. According to the SAA
model, the fragment isotopic and isotonic distributions of
projectile fragmentation should be similar and the difference should
reflect the difference of the proton and neutron distributions. The
shift from the fragment isotopic distribution to the isotonic
distribution in peripheral reactions of $^{40}$Ca reveals the
difference of the proton and neutron density distributions in the
surface region. It should be kept in mind that even in the central
collisions the surface effect of nucleus also exists because in a
specific pipe involved in collisions, the density distribution is
the mixed effect of the surface and the core. But the surface effect
decreases in the central collision since the density of core will
wash out some density difference in the surface region. It make the
shift from the fragment isotopic to isotonic distributions become
smaller. For $^{58}$Ni, the surface effect become unimportant in
most central collisions, which result in the
overlap of the fragment isotopic and isotonic distributions.

For asymmetric neutron-rich nuclei that have a very large $N/Z$
value, more neutrons are pushed to the surface region and there is a
neutron skin structure. In the surface region, the neutron density
is bigger than the proton density and the proton-removing will be
more difficult than the neutron-removing in peripheral reactions.
The wide spread of the fragment isotopic distributions of $^{48}$Ca
and $^{64}$Ni in the top three panels of Fig.~\ref{Ca48Ni64} is the
evidence for that. For $^{48}$Ca and $^{64}$Ni, with $N/Z$ equals
1.4 and 1.286, respectively, there should be a difference in their
neutron density. The similar shapes of the $\Delta N<3$ fragment
isotonic distributions of $^{48}$Ca and $^{64}$Ni reveal the similar
proton density distributions in the surface regions of $^{48}$Ca and
$^{64}$Ni, while the shift on the right sides of the distributions
maybe due to the different diffuseness of $^{48}$Ca and $^{64}$Ni
according to Ref. \cite{Trzc01}. The similarity of the left parts of
the fragment isotopic distributions reveals the similarity of the
neutron density distributions in the surface regions of $^{48}$Ca
and $^{64}$Ni while the shift from the fragment isotopic
distributions of $^{64}$Ni to those of $^{48}$Ca reveals the
difference of the neutron density distribution between them.

\section{summary}
In summary, by investigating the fragment isotopic and isotonic
cross section distributions of 140 $A$ MeV $^{40,48}$Ca + $^9$Be and
$^{58,64}$Ni + $^9$Be reactions, the evidence of isospin effect and
its disappearance in projectile fragmentation has been found. Some
similarity of the fragment isotopic and isotonic distributions
between not only symmetric nuclei but also asymmetric (neutron-rich)
nuclei has been found. These similarities are related to the similar
proton and neutron density distributions in the framework of the SAA
model. These similarities will help to estimate the fragment
production in heavy ion collisions at intermediate energy.

\begin{acknowledgments}
This work was partially supported by the National Science Foundation
of China (Grants 10775168 and 10775039), Shanghai Development
Foundation for Science and Technology (Grant 06QA14062) and the
State Key Program of Basic Research of China (Grant 2007CB815004).
\end{acknowledgments}


\end{document}